\documentclass[aps,preprint,nofootinbib]{revtex4}
\usepackage{amssymb}
\usepackage{amsmath}
\usepackage{bm}
\begin{document}
\title{Inflation as a White Hole explosion from a 5D vacuum}
\author{$^{1, 2}$ Mariano Anabitarte\footnote{anabitar@mdp.edu.ar}, and $^{1, 2}$Mauricio
Bellini\footnote{mbellini@mdp.edu.ar}} \address{$^{1}$
Departamento de F\'{\i}sica, Facultad de Ciencias Exactas y
Naturales, \\
Universidad Nacional de Mar del Plata, Funes 3350, (7600) Mar del
Plata, Argentina. \\ \\
$^{2}$ Instituto de Investigaciones F\'{\i}sicas de Mar del Plata
(IFIMAR), Consejo Nacional de Investigaciones Cient\'{\i}ficas y
T\'ecnicas (CONICET), Argentina.}
\begin{abstract}
Using a new kind of 5D Ricci-flat canonical metric, we obtain by a
static foliation an effective 4D Schwarzschild-de Sitter
hypersurface. We examine some particular initial conditions which
could be responsible for the inflationary expansion of the early
universe, which could be driven by the explosion of a White Hole
(WH). The zeroth order spectrum outside the WH describes quantum
fluctuations, which for a scale invariant power spectrum, can be
expressed in terms of the cosmological constant, or the square
mass of the WH.
\end{abstract}
\maketitle

\section{Motivation}

The Induced Matter Theory (IMT)\cite{IMT} is based on the
assumption that ordinary matter and physical fields that we can
observe in our 4D universe can be geometrically induced from a 5D
Ricci-flat metric with a space-like noncompact extra dimension on
which we define a physical vacuum. The Campbell-Magaard Theorem
(CMT)\cite{campbell} serves as a ladder to go between manifolds
whose dimensionality differs by one. Due to this theorem one can
say that every solution of the 4D Einstein equations with
arbitrary energy momentum tensor can be embedded, at least
locally, in a solution of the 5D Einstein field equations in a
relativistic vacuum: $G_{AB}=0$. For this reason the stress-energy
may be a 4D manifestation of the embedding geometry and therefore,
by making a static foliation on the space-like extra coordinate of
an extended 5D de Sitter spacetime, it is possible to obtain an
effective 4D universe that suffered an exponential accelerated
expansion driven by an effective scalar field with an equation of
state typically dominated by vacuum\cite{ua,ua1,LB,B}. In this
work we shall consider a new kind of 5D Ricci-flat static
canonical metric
\begin{equation}\label{a1}
dS_{(5)}^{2}=F^2(\psi)\,\left[f(r)\,dt^{2}- \frac{dr^2}{f(r)}  -
r^2 \left(d\theta^2 + \sin^2{(\theta)} \,d\phi^2 \right) \right] -
\frac{3}{\Lambda} \left(\frac{dF(\psi)}{d\psi}\right)^2
\,d\psi^{2}.
\end{equation}
Here, $F(\psi)$ is a dimensionless function of the extra
coordinate, $f(r)=1-{2G m \over r} -{\Lambda r^2 \over 3}$ is a
dimensionless function, $\psi$ is the space-like and non-compact
fifth extra coordinate\footnote{In our notation conventions
henceforth, indices $a,b=$ run from 0 to 4, whereas the rest of
indices $i,j,n,l,...=$ run from 1 to 3.}. This is a generalization
of the group of 5D Ricci-flat canonical metrics, that extends a 4D
Swarzschild - de Sitter (SdS) metric. The coordinate $t$ is
time-like and $r,\theta,\phi$ are the usual spherical polar
coordinates. Furthermore, $\Lambda$ is the cosmological constant,
$m$ is the mass of the gravitational source and we shall consider
that the speed of light is dimensionless: $c=1$. The metric
(\ref{a1}) is very interesting because avoids naked singularities
for $m< {1 \over 3 G \sqrt{\Lambda}}$. There cannot be sources
with masses $m>{1 \over 3 G \sqrt{\Lambda}}$, because in these
cases all the roots are complex.

In this work we shall consider the particular case where the mass
takes the particular value $m \equiv {\psi_0 \over 3\sqrt{3} G}={1
\over 3 G \sqrt{\Lambda}}$. If we foliate
$\psi\equiv\psi_0=\sqrt{{3\over \Lambda}}$, the resulting 4D
hypersurface describes a SdS static spacetime with particular
physical properties. The resulting effective 4D hypersurface will
be a spacetime which has a scalar curvature $ ^{(4)} R=
\frac{\epsilon}{4 \Phi^2} \left[ g^{\mu\nu}_{\,\,\,\,\,,4}
\,g_{\mu\nu,4} + \left( g^{\mu\nu}\, g_{\mu\nu,4}
\right)^2\right]$[see \cite{1} and references therein], such that
in our case $\epsilon=-1$ and $\Phi^2 = {\Lambda \over 12
\left(dF/d\psi\right)^2}$. After making the foliation, we obtain
the effective well known 4D SdS metric
\begin{equation}\label{met2}
   dS_{(4)}^{2}=\left[f(r)\,dt^{2}- \frac{dr^2}{f(r)}  - r^2
\left(d\theta^2 + \sin^2{(\theta)} \,d\phi^2 \right) \right].
\end{equation}
For this particular mass value, the horizon is unique and takes
the value\footnote{There are three roots, but two of them are
complex.}: $r_* = \frac{1}{\sqrt{\Lambda}} $. Furthermore, at
$r=r_*$ the effective 4D components of the Ricci-tensor are zero:
$\left.R_{\alpha\beta}\right|_{r_*} =0$, so that at this point the
spacetime is locally flat. But is this a causal horizon? From the
strict point of view at this point there is no causal horizon
because while it is true that $f(r\equiv r_*)=0$, $f(r)$ has the
same sign for any other value of $r\neq r_*$: $f(r\lessgtr r_*)
<0$. Moreover, the bilinear form $dS_{(4)}^2$ has the same sign on
both sides of $r_*$. However, here arises a difference between the
physical properties for $r<r_*$ and $r
>r_*$. For $r< r_*$ the gravitational field (in absence of angular
moment) is attractive, but for $r > r_*$ is repulsive. This means
that {\em outside the false horizon the gravitational field is
always repulsive}. Hence, observers who are at $r>r_*$ can see a
white hole with a false horizon at $r=r_*$.

This can be seen from another point of view: for a massive test
particle outside a spherically symmetric compact object in 5D with
an exterior metric given by (\ref{a1}) the 5D Lagrangian can be
written as $ L_{(5)}=\frac{1}{2}g_{ab}U^{a}U^{b}$. As it is
usually done in literature we can, without loss of generality,
confine the test particles to describe orbits with $\theta=\pi/2$.
Thus, from the Lagrangian $(\ref{b1})$, it can be easily seen that
only $t$ and $\phi$ are cyclic coordinates, so their associated
constants of motion $p_{t}$ and $p_{\phi}$, are
$p_{t}\equiv\frac{\partial\,L_{(5)}}{\partial U^t}$ and
$p_{\phi}\equiv\frac{\partial\,L_{(5)}}{\partial U^{\phi}}$. Since we are dealing with the
case where $f(r)<0$, the
five-velocity condition for a time-like geodesic will be 
\begin{equation}
g_{ab} U^a
U^b=-1. 
\end{equation}
One can use with the constants of motion  $p_t$ and
$p_{\phi}$, and considering $\theta=\pi/2$, we obtain the equation
for the energy conservation. After making a static foliation $\psi
= \sqrt{{3\over \Lambda}}$ [which implies that the observer is
moving with $U^{\psi}=0$ and $F^2(\psi=\sqrt{{3\over
\Lambda}})=1$], can be obtained the effective 4D equation of the
energy for a test particle on the SdS spacetime:
$\frac{1}{2}\left( U^{r}\right)^{2} +  V_{ind}\left(r\right) =
E_{ind}$, with
\begin{eqnarray}
V_{ind}(r) &= & -\frac{1}{2} \left\{ \left[ f(r) +
\frac{\Lambda}{3} r^2 \right] \frac{p^2_{\phi}}{r^2}  -
\left[ f(r) -1\right]\right\}, \\
E_{ind} & = & \frac{1}{2} \left[ p^2_t + \frac{\Lambda}{3}
p^2_{\phi} - 1 \right].
\end{eqnarray}
The interesting thing is that, for $p_{\phi}=0$, (i.e., when the
field is purely gravitational), the Newtonian acceleration $a_N =
-dV_{ind}/dr$, is given by
\begin{equation}
a_N =  - \sqrt{\frac{\Lambda}{3}} \frac{1}{r^2} +
\frac{\Lambda}{3} r,
\end{equation}
and this acceleration is negative for $r<r_*$, but positive for
$r>r_*$, so that for sufficiently large scales the effective
gravitational field becomes repulsive. In the following we shall
explore the cosmological consequences of repulsive gravity in the
early universe.

\section{Inflation from a WH explosion}

As we can see, the metric (\ref{a1}) is written on a static chart
coordinate. In order to get it written on a dynamical chart
coordinate $\lbrace t,r,\theta,\phi \rbrace$, let us use the
coordinate transformation\cite{planar} given by
\begin{equation}\label{a2}
R=a\,r\left[1+\frac{1}{6ar \sqrt{\Lambda}}\right]^{2},\quad
T=t+\sqrt{\frac{\Lambda}{3}}\int^{r}dR\,\frac{R}{f(R)}\left(1-\frac{2}{3
\sqrt{\Lambda} R}\right)^{-1/2}, \psi=\psi,
\end{equation}
where $a(t)=e^{\sqrt{\frac{\Lambda}{3}}t}$ is the scale factor,
and $\sqrt{\frac{\Lambda}{3}}$ is the Hubble constant. Thus the
line element (\ref{a1}) can be written in terms of the conformal
time $\tau$ as
\begin{equation}\label{a3}
dS_{(5)}^{2}=\left[F(\psi)\right]^{2}\left[U(\tau,r)d\tau^{2}-V(\tau,r)\left(dr^{2}+r^{2}(d\theta^{2}
+sin^{2}\theta d\phi^{2})\right)\right]-\frac{3}{\Lambda}
\left(\frac{dF(\psi)}{d\psi}\right)^2 d\psi^{2},
\end{equation}
where the metric functions $U(\tau,r)$ and $V(\tau,r)$ are given
by
\begin{equation}\label{a4}
U(\tau,r)=a^{2}(\tau)\left[1-\frac{1}{6ar
\sqrt{\Lambda}}\right]^{2}\left[1 +\frac{1}{6ar
\sqrt{\Lambda}}\right]^{-2},\quad V(\tau,r)
=a^{2}(\tau)\left[1+\frac{1}{6ar \sqrt{\Lambda}}\right]^{4},
\end{equation}
with $d\tau=a^{-1}(\tau)dt$ and
$a(\tau)=-1/(\sqrt{\frac{\Lambda}{3}}\tau)$, so that the Hubble
parameter is a constant given by
$\sqrt{\frac{\Lambda}{3}}=a^{-2}\, \frac{d a}{d\tau}$.

Now we consider a 5D massless scalar field which is free of any
interactions. The dynamics is given by a Klein-Gordon equation:
$\frac{1}{\sqrt{|g_{(5)}|}}\frac{\partial}{\partial
y^{a}}\left[\sqrt{|g_{(5)}|}g^{ab}\varphi_{,b}\right]=0$, where
$\sqrt{|g_{(5)}|}=F^4(\psi) \sqrt{{3\over \Lambda}} \left|
dF/d\psi\right| U^{1/2}V^{3/2}r^{2}sin\theta$ is the determinant
of the covariant metric tensor $g_{ab}$. It can be demonstrated
that $\varphi(\tau,r,\theta,\phi,\psi)$ can be separated in the
form $\varphi(\tau,r,\theta,\phi,\psi)\sim
\Phi(\tau,r)G(\theta,\phi)\Omega(\psi)$, such that the
differential equation for $\Omega(\psi)$ is
\begin{equation}
\Omega'' + \left[ \frac{4
\left(\Lambda/3\right)^2}{\left(F'\right)^3 F} - \frac{F''}{F'}
\right] \Omega' - \left( \frac{\Lambda M^2}{3 (f')^2 f^2} \right)
\Omega(\psi) = 0,
\end{equation}
where $M^2$ is a separation constant and the prime denotes the
derivative with respect to $\psi$.

\subsection{The 4D induced field equation}

Assuming that the 5D spacetime can be foliated by the hypersurface
$\psi=\sqrt{{3\over \Lambda}}$, from the metric (\ref{a3}) we
obtain that the 4D induced metric is given by
\begin{equation}\label{b1}
dS_{(4)}^{2}=U(\tau,r)d\tau^{2}-V(\tau,r)[dr^{2}+r^{2}(d\theta^{2}+\sin^{2}\theta
d\phi^{2})],
\end{equation}
where the metric functions $U(\tau,r)$ and $V(\tau,r)$ are given
by (\ref{a4}). We shall consider that
${\bar\varphi}(\tau,r,\theta,\phi)=\varphi(\tau,r,\theta,\phi,\psi_0)$
is the effective scalar field induced on the effective 4D
hypersurface obtained after making the foliation. Furthermore, for
consistency we shall require that $F\left(\psi=\sqrt{{3\over
\Lambda}}\right)=1$, in order to make
$\left.\sqrt{|g_{(5)}|}\right|_{\psi=\sqrt{{3\over
\Lambda}}}=\sqrt{|g_{(4)}|}$, where $g_{(4)}$ is the determinant
of the metric (\ref{b1}). We expand the induced scalar field
${\bar\varphi}$ as\cite{egr}
\begin{equation}
\bar\varphi(\vec r,\tau) = \int^{\infty}_{0} dk\,\sum_{lm}
\left[a_{klm} \bar\Phi_{klm}(\vec r,\tau)+a^{\dagger}_{klm}
\bar\Phi^*_{klm}(\vec r,\tau)\right],
\end{equation}
where $\bar\Phi_{klm}(\vec r,\tau)= k^2 \,j_l\left(kr\right)
\bar\Phi_{kl}(\tau) Y_{lm}(\theta,\phi)$, such that
$Y_{lm}(\theta,\phi)$ are the spherical harmonics, $j_{l}(kr)$ are
the spherical Bessel functions and the annihilation and creation
operators obey the algebra $\left[a_{klm},
a^{\dagger}_{k'l'm'}\right] = \delta(k- k') \delta_{ll'} \delta_{m
m'}$, $\left[a_{klm}, a_{k'l'm'}\right]=\left[a^{\dagger}_{klm},
a^{\dagger}_{k'l'm'}\right]=0$. The mean square fluctuations can
be obtained by using the theorem of the spherical harmonics
\begin{equation}\label{fluct}
\left< E\left| \bar\varphi^2\left(\vec r,\tau\right)
\right|E\right> = \int^{\infty}_{0} \frac{dk}{k} \sum_{l}
\frac{2l+1}{4\pi} k^5 j^2_l(kr)
\left|\bar\Phi_{kl}(\tau)\right|^2,
\end{equation}
where $\left|E\right>$ is an arbitrary quantum state. We shall
assume that\footnote{The equation for $\bar\Phi(r,\tau)$ on the
effective 4D hypersurface described by (\ref{b1}), and can be
written as
\begin{eqnarray}
\frac{\partial^2 {\bar\Phi_l}}{\partial\tau^2} & - &
\frac{2}{\tau} \frac{\partial {\bar\Phi_l}}{\partial\tau} -
\frac{2}{r} \frac{\partial {\bar\Phi_l}}{\partial r} -
\frac{\partial^2 {\bar\Phi_l}}{\partial r^2} - \left[\frac{l\left(
l+1\right)}{r^2} - M^2\right] {\bar\Phi_l}  \nonumber  \\
& = & \left( 1- \frac{V}{U}\right) \frac{\partial^2
{\bar\Phi_l}}{\partial\tau^2} - \left[ \frac{2}{\tau} +
\frac{1}{\sqrt{U V}} \frac{\partial}{\partial\tau}
\left(\frac{V^3}{U}\right)^{1/2} \right] \frac{\partial
{\bar\Phi_l}}{\partial\ \tau}  -  M^2 \left(V -1\right)
{\bar\Phi_l} + \frac{1}{2} \left( \frac{1}{U} \frac{\partial
U}{\partial r} + \frac{1}{V} \frac{\partial V}{\partial r}\right)
\frac{\partial\bar\Phi_l}{\partial r}. \nonumber
\end{eqnarray}}
$\bar{\varphi}(\tau,r,\theta,\phi)=\bar{\Phi}(\tau,r)\bar{G}(\theta,\phi)$.
Next,using the fact that $\epsilon$ is a small  we propose the
following expansion for $\bar\Phi_l $ in orders of $\epsilon$:
$\bar\Phi_l(r,\tau) = \bar\Phi^{(0)}_l + \bar\Phi^{(1)}_l + ....$.
Thus, we are now able to calculate solutions for
$\bar{\Phi}_l(r,\tau)$ at zeroth order in the expansion. The
spectrum for the square fluctuations (\ref{fluct}) can be
calculated at zeroth order
\begin{equation}
{\cal P}_k(\tau) =  \frac{k^3}{2\pi^2}
\left|\bar\Phi^{(0)}_{kl}\right|^2 + ...\,, \label{spect}
\end{equation}

\subsection{Scalar field dynamics and zeroth order power spectrum outside the WH}

The dynamics for a massive scalar field in a spatially homogeneous
de Sitter case is described by
\begin{equation}
\frac{\partial^2 {\bar\Phi^{(0)}}_l}{\partial\tau^2} -
\frac{2}{\tau} \frac{\partial {\bar\Phi}^{(0)}_l}{\partial\tau} -
\frac{2}{r} \frac{\partial {\bar\Phi}^{(0)}_l}{\partial r} -
\frac{\partial^2 {\bar\Phi}^{(0)}_l}{\partial r^2} -
\left[\frac{l\left( l+1\right)}{r^2} - M^2 a^2(\tau)\right]
{\bar\Phi}^{(0)}_l=0,
\end{equation}
where the last term corresponds to the induced mass of the scalar
field. If we use the Bessel transform: $ \bar\Phi^{(0)}_l(r,\tau)
=  \int^{\infty}_{0} dk\, k^2\, j_l(kr)\,
{\bar\Phi^{(0)}}_{kl}(\tau)$, we obtain the zeroth order dynamics
for the modes ${\bar\Phi^{(0)}}_{kl}$
\begin{equation}
\frac{\partial^2 {\bar\Phi^{(0)}}_{kl}}{\partial\tau^2} -
\frac{2}{\tau} \frac{\partial {\bar\Phi^{(0)}}_{kl}}{\partial\tau}
+ \left(k^2 + \frac{3 M^2}{\Lambda \tau^2}\right)
{\bar\Phi^{(0)}}_{kl}=0.
\end{equation}
If we use the Bunch-Davies vacuum, we obtain the normalized modes
solution: ${\bar\Phi^{(0)}}_{kl} = A_1 \, \left(-\tau\right)^{3/2}
{\cal H}^{(1)}_{\nu}\left[-k\,\tau\right] + A_2 \,
\left(-\tau\right)^{3/2} {\cal
H}^{(2)}_{\nu}\left[-k\,\tau\right]$, where ${\cal
H}^{(1,2)}_{\nu}$ are respectively the first and second kind
Hankel functions with $\nu^2 = {9 \over 4} - {M^2\over H^2}$. The
normalization constants are $A_2 = -\frac{\sqrt{\pi}
\sqrt{\Lambda}}{2 \sqrt{3}}\,e^{-i \nu \pi/2}$, $A_1 = 0$. We are
interested in the spectrum of the fluctuations without sources,
such that $l=0$. The spectrum is
\begin{equation}
{\cal P}^{(0)}_{kl}(\tau) = \frac{k^3}{2\pi^2}
\left|\bar\Phi^{(0)}_{kl} \right|^2 = \frac{\Lambda}{3\pi}
\left(\frac{-k \tau}{2}\right)^{3}  \, {\cal
H}^{(2)}_{\nu}[-k\tau] \, {\cal H}^{(1)}_{\nu}[-k\tau],
\end{equation}
where $k$ given by $k = \frac{2\pi}{a(\tau) r} \left[1+\frac{1}{2
\sqrt{3} \sqrt{\Lambda} a(\tau) r}\right]^{-2}$, is the wave
number on a physical frame.

Outside the WH the fluctuations are small and the wavenumber of
these fluctuations on scales of size ${\sqrt{3} \over
\sqrt{\Lambda}}$, is
\begin{equation}\label{aa1}
k_H \simeq \frac{2\pi}{a(\tau) r_H} = -\frac{2\pi
\sqrt{\Lambda}\tau}{\sqrt{3} r_H},
\end{equation}
where we have made use of the fact that $a(\tau)=
-\sqrt{3}/(\sqrt{\Lambda}\tau)$, with $\tau \leq 0$. Furthermore,
at the end of inflation $\tau \rightarrow 0$, so that ${\cal
H}^{(2)}_{\nu}[-k\tau] \simeq {-i\over \pi} \Gamma(\nu)
\left(-k\tau/2\right)^{-\nu}$. The power spectrum on scales close
to the Hubble horizon is
\begin{equation}
\left.{\cal P}^{(0)}_{kl}(\tau)\right|_{H} = \frac{k^3}{2\pi^2}
\left|\bar\Phi^{(0)}_{kl}\right|^2_{H} \simeq \left(\frac{\pi
\sqrt{\Lambda}\tau^2}{\sqrt{3} r_H}\right)^{3-2\nu}
\frac{\Gamma^2(\nu) \Lambda}{3 \pi^3},
\end{equation}
which depends on the mass of the inflaton field $M$, because
$\nu^2 = {9\over 4} - {3M^2\over \Lambda}$. For a nearly  scale
invariant power spectrum: $\nu \simeq 3/2$, we obtain
\begin{equation}
\left.{\cal P}^{(0)}_{kl}(\tau)\right|_{H,\nu \simeq 3/2} \simeq
\frac{\Lambda}{12\pi^2},
\end{equation}
which describe quantum fluctuations. Notice the dependence of this
spectrum with the cosmological constant (or the square mass of the
WH).

\section{Final Comments}

Using a new kind of canonical metric here proposed, we have
explored the possibility that the inflationary early universe
could be driven by the explosion of a WH with a mass which is
related to the value of the foliation of the non-compact extra
dimension [$\psi=\psi_0$], and hence of the cosmological constant:
$m \equiv {\psi_0 \over 3\sqrt{3} G}={1 \over 3 G
\sqrt{\Lambda}}$. This idea appears to be very attractive because
excludes the problem of the initial singularity taking into
account the idea of a large-scale repulsive gravitational field.

\section*{Acknowledgements}
\noindent The authors acknowledge UNMdP and CONICET Argentina for
financial support.

\end{document}